%Paper: hep-th/9510228
%From: Matt Strassler <strasslr@physics.rutgers.edu>
%Date: Tue, 31 Oct 1995 00:09:47 -0500

\input harvmac
%\draftmode
\def\np#1#2#3{Nucl. Phys. B{#1} (#2) #3}
\def\pl#1#2#3{Phys. Lett. {#1}B (#2) #3}

\def\physrev#1#2#3{Phys. Rev. {D#1} (#2) #3}

\def\prep#1#2#3{Phys. Rep. {#1} (#2) #3}

\def\ev#1{\langle#1\rangle}

\def\tilde{\widetilde}
\def\frac#1#2{{#1\over#2}}
\def\half{\frac{1}{2}}
\def\al{\alpha}
\def\bt{\beta}

\def\psqr#1#2{{\vcenter{\vbox{\hrule height.#2pt
	\hbox{\vrule width.#2pt height#1pt \kern#1pt
	\vrule width.#2pt}
	\hrule height.#2pt \hrule height.#2pt
	\hbox{\vrule width.#2pt height#1pt \kern#1pt
	\vrule width.#2pt}
	\hrule height.#2pt}}}}
\def\sqr#1#2{{\vcenter{\vbox{\hrule height.#2pt
	\hbox{\vrule width.#2pt height#1pt \kern#1pt
	\vrule width.#2pt}
	\hrule height.#2pt}}}}

\Title{hep-th/9510228, RU-95-67}
{\vbox{\centerline{A Chiral $SU(N)$ Gauge Theory}
\centerline{}\centerline{and its Non-Chiral $Spin(8)$ Dual}}}
\bigskip
\centerline{P. Pouliot and M.J. Strassler}
\vglue .5cm
\centerline{Department of Physics and Astronomy}
\centerline{Rutgers University}
\centerline{Piscataway, NJ 08855-0849, USA}

\bigskip

\noindent

We study supersymmetric $SU(N-4)$ gauge theories with a symmetric tensor
and $N$ antifundamental representations.  The theory with $W=0$ has
a dual description in terms of a non-chiral $Spin(8)$ theory with one
spinor and $N$ vectors. This duality flows to the $SO(N)$ duality of Seiberg
and to a duality proposed by one of us.
It also flows to dualities for a number of $Spin(m)$ theories, $m\le 8$.
For $N=6$, when an ${\cal N}=2$ SUSY superpotential is added,
the singularities of Seiberg and Witten are recovered. For $N\le 6$,
a mass for the spinor generates the branches of $SO(8)$ theories
found by Intriligator and Seiberg.  Other phenomena include
a classical constraint mapped to an anomaly equation under duality
and an intricate consistency check on the renormalization group flow.

\Date{10/95}

\nref\powerh{N. Seiberg, {\it The Power of Holomorphy -- Exact
Results in 4D SUSY Field Theories}, Proc. of PASCOS 94, hep-th/9408013;
{\it The Power of Duality
 -- Exact Results in 4D SUSY Field Theories},
Proc. of PASCOS 95 and Proc. of the
Oskar Klein Lectures, hep-th/9506077, RU-95-37, IASSNS-HEP-95/46}
\nref\kinsrev{K. Intriligator and N. Seiberg,
hep-th/9509066, RU-95-48, IASSNS-HEP-95/70}
\nref\ads{I. Affleck, M. Dine and N. Seiberg, \np{241}{1984}{493};
\np{256}{1985}{557}}%
\nref\rus{A.I. Vainshtein, V.I. Zakharov and M.A. Shifman,
Sov. Phys. Usp. 28 (1985) 709;
M.A. Shifman and A. I. Vainshtein, \np{277}{1986}{456};
\np{359}{1991}{571}}
%\nref\sv{M.A. Shifman and A. I. Vainshtein, \np{277}{1986}{456};
%\np{359}{1991}{571}}
\nref\cern{D. Amati, K. Konishi, Y. Meurice, G.C. Rossi and G.
Veneziano, \prep{162}{1988}{169} }
\nref\sem{N. Seiberg, \np{435}{1995}{129}, hep-th/9411149 }%
\nref\son{K. Intriligator and N. Seiberg, \np{444}{1995}{125};
hep-th/9506084, RU-95-40, IASSNS-HEP-95/48}
\nref\emop{R.G. Leigh
and M.J. Strassler, \np{447}{1995}{95}, hep-th/9503121}
\nref\olive{C. Montonen and D. Olive, \pl{72}{1977}{117}}
\nref\nsewone{N. Seiberg and E. Witten, \np{426}{1994}{19}, hep-th/9408013}
\nref\nsewtwo{N. Seiberg and E. Witten,  \np{431}{1994}{484}, hep-th/9407087}
\nref\kut{D. Kutasov, \pl{351}{1995}{230}, hep-th/9503086}
\nref\aharony{O. Aharony, J. Sonnenschein and S. Yankielowicz,
hep-th/9504113, TAUP-2246-95, CERN-TH/95-91}
\nref\kuschwim{D. Kutasov and A. Schwimmer,
\pl{354}{1995}{315}, hep-th/9505004}
\nref\intpou{K. Intriligator and P. Pouliot, hep-th/9505006,
\pl{353}{1995}{471}}
\nref\berk{M. Berkooz, hep-th/9505067, RU-95-29}
%\nref\intril{K. Intriligator, \np{448}{1995}{187}, hep-th/9505051}
%\nref\leighstr{R.G. Leigh
%and M.J. Strassler, \pl{356}{1995}{492}, hep-th/9505088}
\nref\ilstr{K. Intriligator, \np{448}{1995}{187}, hep-th/9505051;
R.G. Leigh and M.J. Strassler, \pl{356}{1995}{492}, hep-th/9505088;
K. Intriligator, R.G. Leigh and M.J. Strassler,
hep-th/9506148, RU-95-38}
\nref\spinseven{P. Pouliot,  \pl{359}{1995}{108}, hep-th/9507018}
\nref\suantisym{P. Pouliot, hep-th/9510148, RU-95-66}
\nref\spinten{P. Pouliot and M. Strassler, in preparation}
\nref\KonishiPS{T.E. Clark, O. Piguet and K. Sibold,
Ann. Phys. 109 (1977) 418; \np{143}{1978}{445}; \np{159}{1}{1979};
S.J. Gates, M.T. Grisaru, M. Rocek and W.
          Siegel, \it Superspace\rm, Benjamin, Reading, MA 1983;
K. Konishi, \pl{135}{1984}{439}}
\nref\ils{K. Intriligator, R.G. Leigh and N. Seiberg,
\physrev{50}{1994}{1092}, hep-th/9403198;
K. Intriligator, \pl {336}{1994}{409}, hep-th/9407106}

With the pioneering work of Seiberg,
it has become clear that ${\cal N}=1$ supersymmetry renders
possible the precise study of dynamical phenomena in
four-dimensional quantum field theory.
For recent reviews and lists of
references, see \refs{\powerh,\kinsrev};
for earlier work, see \refs{\ads - \cern}.
Crucial for understanding dynamics of ${\cal N}=1$ SUSY
theories is duality \sem, a generalization
\refs{\sem - \emop} of the Montonen-Olive
duality \refs{\olive-\nsewtwo} of extended supersymmetry.
One or more duals of many theories have
been found \refs{\sem, \son, \nsewone - \spinten},
but the general rules are not yet
understood and for most theories no dual representation is known.

In this letter, we generalize the examples found in \spinseven.
There, an $SU(N-4)$ gauge theory, with a field $S$ in the
symmetric tensor representation
and $N$ fields $Q$ in the antifundamental representation, and with
superpotential $W=\det S$, was argued to have
a dual description using $Spin(7)$ with spinors.
In this letter, we suggest that the same $SU(N-4)$ gauge theory
with $W=0$ has a dual description in terms
of a $Spin(8)$ gauge theory with vectors and one spinor.
We give a number of consistency arguments that strongly support this claim.

As we will show, this pair of theories connects the dualities found in
\sem\ to the one found in \spinseven.  We also can derive duals
for the many theories lying along the flat directions of the
$Spin(8)$ theory.  There are a number of interesting phenomena, including
a classical constraint mapped to an anomaly equation, a theory
which flows to a particular ${\cal N}=2$ duality found in \nsewtwo,  the
emergence of branches when a spinor of $Spin(8)$ becomes massive,
and a complex interplay of flavor symmetries which, as in \nsewtwo,
are represented very differently in the electric and magnetic theories.

\newsec{$SU(N-4)$ with superpotential $W=0$}
\subsec{The electric theory}

Consider an ${\cal N}=1$ supersymmetric $SU(N-4)$ gauge theory with a
field $S$ in the symmetric tensor representation and $N$ fields
$Q^i$ in the antifundamental representation.
The global symmetry is $SU(N)\times U(1) \times U(1)_R$,
where $U(1)_R$ is an R-symmetry, under which $S$, $Q^i$ transform as
$\left({\bf 1},\ -2N,\ 12R_0\right)$,
$\left({\bf N},\ 2N-4,\ 6[N-5]R_0\right)$,
where $R_0^{-1}=(N+1)(N-4)$.   Note that there is a one-parameter
family of anomaly-free $R$ symmetries;
our particular choice is for later convenience.
There are no additional discrete symmetries.
Any discrete symmetry can be redefined, using the $U(1)$
symmetries, so that it acts only on the fields $Q$;
but then it must be a subgroup of $Z_N$, which itself
is already the center of the $SU(N)$ flavor symmetry.

The D-term potential for the scalar components of $S$, $Q^i$ has many
flat directions. Up to flavor and gauge symmetry rotations, these are
\eqn\adssolution{
Q_\alpha^i = \pmatrix{v_1 & & & &|&\cr
& v_2 & & &|&\cr
& & \cdots& &|&0\ 0\ 0\ 0 \cr
& & & v_{N-4} & |&  \cr} \qquad {\rm and}\qquad
S^{\al\bt} = {\rm diag}(w_1,\dots,w_{N-4})}
where $|v_k|^2-2|w_k|^2$ is independent of $k$.
The independent gauge invariant operators of this theory
are mesons $M^{ij}=Q^i S Q^j$  which are symmetric in their
flavor indices, a flavor singlet $U=\det S$, and a
number of baryons, totally antisymmetric in flavor,
namely $B\equiv Q^{N-4}$ contracted with one $SU(N-4)$ epsilon
tensor, and $B'_n=Q^{N-4-2n}S^{N-4-n} W^n_\alpha \ (n=0,1,2)$
contracted with two $SU(N-4)$ epsilon tensors.  Classically
these operators satisfy some constraints; for
example, $M^{N-4} = UB^2$, where the indices of the $N-4$ factors
of $M$ are completely antisymmetrized.  Another constraint
is that $B'_0=UB$.

Holomorphy and the symmetries forbid any dynamically generated
superpotential. Note that $\det M$ and the $SU(N)$ singlet
$M^4B^2$ vanish identically. The only non-zero flavor singlet
is $\det S$, and since it is charged
under the $U(1)$, no invariant superpotential can be written.
The quantum moduli space is therefore the same as the
classical one.

Those operators not containing $W_\alpha$ have flat directions
associated with them.
If $\vev{M}$ has rank $k$, then it breaks the gauge symmetry
to $SU(N-4-k)$ with a symmetric tensor $\hat S$ and $N-k$
antifundamentals; for $k=N-5$, the gauge group
is broken and there are massless
Goldstone bosons, Higgs bosons, and five singlets $Q_i$, while for
$k=N-6$ only Goldstone and Higgs multiplets are present.  The scales
of the high- and low-energy theory are related by
\eqn\Mscale{
\Lambda_{SU(N-4)}^{2N-11}\propto
\left(\prod_{i=1}^{k} \vev{Q^i}\right)^2 \Lambda_{SU(N-k-4)}^{2(N-k)-11}}
where the D-term constraint $\vev{S^{ii}}\propto\vev{Q^i}$ has been used.
(We have not computed the numerical threshhold factors in this and similar
relations.) A vacuum expectation value for
the operator $U$ breaks the theory to $SO(N-4)$ with
$N$ fields in the vector representation remaining; in this
case the scales are related by\foot{We thank K. Intriligator and R. Leigh
for a discussion of this scaling relation.}
\eqn\Uscale{
\Big[\Lambda^{2N-11}_{SU(N-4)}
\Big]^2\propto\vev{S}^{2N-4}\Lambda^{2N-18}_{SO(N-4)}.}
The scale of the $SU$ group is squared since two of its instantons
are needed here to make an $SO$ instanton.
Along the $\ev{B}$ flat direction the theory is completely broken.

All of these theories (for $N\geq 6$) are asymptotically free.
We will present evidence that for $6\leq N\leq 16$ the theory
flows to an interacting fixed point, while for $N>16$ the theory flows to
a free fixed point. Note also that for $N=6$ the theory has the matter
content of an ${\cal N}=2$ theory studied in \nsewtwo\ but without the
${\cal N}=2$ superpotential.  We will discuss the relation with
the known ${\cal N}=2$ theory below.

\subsec{The ``magnetic'' $Spin(8)$ theory}

The magnetic dual of this theory is a $Spin(8)$ gauge
theory with $N$ vector representations $q_i$, a spinor
representation $p$, and $Spin(8)$ singlets $M^{ij}$ and
$U$.\foot{The $Spin(8)$ group has dimension $28$ and
has three representations of dimension
$8$, known as $8_v$, $8_s$, and $8_c$: the vector, spinor
and conjugate spinor
representations.  These are permuted
under the $S_3$ triality symmetry of the group;
only relative labellings are meaningful.}
Its  classical superpotential is $\tilde W=Mqq/\mu_1^2+Upp/\mu_2^{N-5}$.
It implements the constraints $q_iq_j=pp=0$ in the infrared.  The scales
$\mu_1,\mu_2$ are needed for dimensional consistency under the duality
transformation.  The relationship between the scales $\Lambda$
of the $SU(N-4)$ theory and $\tilde \Lambda$ of the $Spin(8)$
theory is\foot{We thank D.~Kutasov and A.~Schwimmer for a discussion
of the scales.}
\eqn\scales{
\Big[\Lambda^{2N-11}\Big]^2\tilde
\Lambda^{17-N}\propto\mu_1^{2N}\mu_2^{N-5}.}
For simplicity we will drop all factors of $\mu_1,\mu_2$.

Under the $SU(N)\times U(1) \times U(1)_R$ global symmetries,
the fields
$q$, $p$, $M$, $U$ transform as
$({\bf \overline{ N}},\ 4-N,\ 1-\tilde R_0)$,
$({\bf 1},\ N[N-4],\ 1-\tilde R_0)$,
$({\bf \half N[N+1]},\ 2N-8,\ 2\tilde R_0)$,
$({\bf 1},\ -2N[N-4],\ 2\tilde R_0)$, where $\tilde R_0 = {6\over N+1}$.
The symmetries, holomorphy, and smoothness near
the origin $M, U, q, p=0$ uniquely determine the magnetic
dual superpotential.  (For $N=6,7$ other operators
may appear.)

The independent gauge invariant operators of this $Spin(8)$ theory
are the fundamental singlets $M^{ij}$ and $U$, the mesons $q^2$ and
$p^2$ (which are redundant
as a result of the superpotential), the baryons
$b'_n = q^{8-n}\tilde W_\alpha^n$ contracted with a
$Spin(8)$ epsilon tensor, and $b=q^4p^2$, where the
vectors $q_i$ are combined antisymmetrically into a
${\bf 70}={\bf 35_s}+{\bf 35_c}$ representation of $Spin(8)$ and the
spinors $p$ are combined symmetrically into a ${\bf 35_s}$.

For $N\ge 17$, the magnetic theory is not asymptotically
free, so it flows to a free theory of gluons and quarks
in the infrared. Accordingly, the electric description is not
valid there.  For $N\geq 17$,
there is no $R$ symmetry for which both $QSQ$ and $\det S$ have
charges greater than $2/3$, so an interacting conformal
field theory involving the fields $Q^i$, $S$ cannot be unitary.  Thus,
the $SU(N-4)$ theories described above
are actually free $Spin(8)$ theories in the infrared for $N\ge 17$.

\subsec{Consistency Checks on the Duality}

In the following, we present a
number of consistency checks on the $SU(N-4)$/$Spin(8)$ duality.
First, the 't Hooft anomaly matching conditions are satisfied.  Next,
there is a correspondence, which preserves the global symmetries,
between the gauge invariant operators of the electric and magnetic theories.
\eqn\opermap{\matrix{
SU(N-4)\ \ theory      &\to           & Spin(8)\ dual\ theory &\cr
       \cr
QSQ                  & \to & M       & \cr
\det S               & \to & U       &  \cr
B=Q^{N-4}  & \to & b\equiv q^4p^2 & \cr
B'_n=Q^{N-4-2n}S^{N-4-n} W^n_\alpha  & \to &
                     b'_n\equiv q^{4+2n} \tilde W^{2-n}_\alpha
&(n=0,1,2)  .   \cr } }

The operator $B'_0$ satisfies the classical constraint
$B'_0=UB$ in the $SU(N-4)$ theory.  The operator $b'_0=q^4\tilde W_{\alpha}^2$
is similarly constrained by a quantum mechanical effect, which
can most easily be seen along one of the flat directions of $Spin(8)$.
If we add $mM$ to the superpotential, where $m$ is rank 4, this leads to
an expectation value $\vev{q_iq_j}\propto m$ of rank 4,
breaking the magnetic theory to $Spin(4)\approx SU(2)\times SU(2)$.  The
spinor $p$ splits into two spinors $p_a$ of the first $SU(2)$
and two spinors $\bar p_{\dot a}$ of the second.
We have two independent $W_\al^{(i)}$, $i=1,2$, one for each
$SU(2)$.  The chiral anomaly \KonishiPS\
states that $\phi\ (\del W/\del \phi)\sim W_\al^2$
for a chiral superfield $\phi$ charged under a gauge group with
field strength $W_\al^2$.  Putting this all together,
we have $Mqq  \propto W_\al^{(1)2}+W_\al^{(2)2}$,
$Up_1p_2\propto W_\al^{(1)2}$,
$U\bar p_{\dot 1}\bar p_{\dot 2}\propto W_\al^{(2)2}$.
The field $W_\al^{(1)2}+W_\al^{(2)2}$
is massive, so the equations $Mqq = U(pp+\bar p\bar p) =0$ hold
in the low-energy theory (unless there is gaugino condensation.)
However, the other linear combination $W_\al^{(1)2}-W_\al^{(2)2}$ is
massless \refs{\ils,\sem,\son}, so there is an operator equation
$U(p_1p_2-\bar p_{\dot 1}\bar p_{\dot 2})\propto W_\al^{(1)2}-W_\al^{(2)2}$
involving the light fields of the $Spin(4)$ theory.  When we
set $m$ back to zero, thereby returning to $Spin(8)$, we see
this anomaly equation implies the operator relation $b_0'\propto Ub$,
in correspondence with the classical relation $B_0'=UB$.
Related phenomena were observed in \son.

Another check on the duality is that it is connected to
other known dualities by renormalization group flow.
If we add the operator $U=\det S$ to the electric
superpotential, the theory is that studied in \spinseven,
which was shown to be dual to $Spin(7)$ with $N$ spinors.
In the magnetic theory, the superpotential $W=U+Mqq+Upp$
causes $\vev{pp}$ to be nonzero, breaking $Spin(8)$ to
$Spin(7)$ and turning the $N$ vectors $q$ of $Spin(8)$
into $N$ eight-dimensional spinors of $Spin(7)$, with
superpotential $W=Mqq$.  On the other hand, if we
go along a flat direction where $\vev{U}\neq 0$, then
this breaks $SU(N-4)$ to $SO(N-4)$ with $N$ vectors,
which is dual \refs{\sem,\son} to $SO(8)$
with $N$ vectors. In the magnetic theory, an expectation
value for $U$ gives mass to the spinor $p$, leaving
$SO(8)$ with $N$ vectors and a superpotential $W=Mqq$.
It can be checked in both cases that the operator map
proposed in equation \opermap\ flows to the correct operator
map in the dual theory.

Now consider the flat directions along which $\vev{M}$ has
rank $k$. In this
paragraph will be using dynamical results derived later
for the $Spin(8)$ theory. If $k<N-5$, the electric theory
flows to a theory of the same type with $N-4-k$ colors;
in the magnetic theory, $k$ vectors become massive,
leaving $Spin(8)$ with $N-k$ vectors and one spinor,
and thus preserving the duality.  If $k\geq N-5$ the situation
is more subtle, though in the end it is similar
to \sem.   For $k=N-5$ the electric gauge group
is completely broken and five singlet quarks $Q^i$
remain as unconstrained massless degrees of freedom.  The magnetic
theory of $Spin(8)$ with five vectors and a spinor confines;
the confined theory has $M^{ij},U,N_{ij}=q_iq_j$, $T=pp$, and
$b^i=\epsilon^{ijklm}q_jq_kq_lq_mpp$
($i,j,\dots=1,\dots,5$) as its massless spectrum.
There is a classical constraint $T^2\det N+Nbb=0$ which is modified
by quantum effects to $T^2\det N+Nbb=T\tilde \Lambda_L^{12}$,
where $\Lambda_L$, the scale of the low-energy $Spin(8)$ theory,
is related to that of the high-energy theory by
$\tilde\Lambda_L^{12}= \tilde\Lambda_H^{17-N}M^{N-5}$.
Since the superpotential $W=MN+ UT +X(T\det N+Nbb/T-\tilde\Lambda_L^{12})$
sets $N,T$ to zero, the $b^i$ are actually unconstrained.  As in \sem,
these composite baryons are to be identified as the five singlet
quarks $Q^i$ of the electric theory.
If $k=N-4$, then the classical constraint
$M^{N-4} = U B^2$ must be obeyed in the electric theory.
The magnetic $Spin(8)$ theory confines and generates a
non-perturbative superpotential, so that
$\tilde W = MN+UT-T\tilde\Lambda_L^{13}/(T^2\det N+b^2)$.  The
equations for $M,U$ set $N_{ij}=T=0$, while the
equation for $T$ is
$U= \tilde\Lambda_L^{13}/b^2= \tilde\Lambda_H^{17-N}M^{N-4}/b^2$,
which agrees with the electric constraint.
The electric theory cannot have $k>N-4$.
Correspondingly, the magnetic
theory, with a
dynamical superpotential proportional to
$[\tilde\Lambda^{17+k-N}_L/T\det N]^{1/(5-N+k)}$,
has no ground state.

\newsec{$W = y_{ij}M^{ij}$; More Dual Pairs}

Next, consider adding the terms $y_{ij}M^{ij}$ to
the superpotential, where the rank of $y$ is $k$,
and study the dual theories obtained under the flow.
The magnetic theory, with $\tilde W = M(y+qq)+Upp$,
breaks to $Spin(8-k)$
with $N-k$ vectors and enough spinor representations
$p_a (\bar p_{\dot a})$ to make up an $8_s$ of $Spin(8)$.
The low-energy superpotential is $\tilde W = Mqq+U(pp)$,
where $(pp)$ is a mass operator for all the low-energy spinors.

These theories are consistent with other examples of duality.
For example, if we give an expectation value to the operator $U$,
the electric theory breaks to $SO(N-4)$,
and the superpotential $W=y_{ij}Q^i\vev{S}Q^j$ gives
mass to $k$ fields $Q$, leaving $N-k$ vector representations.
This theory is dual to $SO(8-k)$ with $N-k$ vectors,
which is indeed what remains in the magnetic theory \refs{\sem,\son},
since $\vev{U}$ gives mass to all of the spinors.
Next, consider the effect of adding $U=\det S$ to the superpotential,
which causes the spinors of the magnetic theory to condense.  As
an example, if $k=1$, the magnetic theory is
$Spin(7)$ with $N-1$ vectors and one spinor; when
$U$ is added to $\tilde W$, the spinor condenses
and breaks the theory to $G_2$ with $N-1$ fundamentals.
This is dual to $SU(N-4)$ with fields $W=\det S + Q^1 S Q^1$,
in agreement with \spinseven.

When $k\geq 7$, the magnetic theory is completely
broken to a theory of singlets.  Duality implies
that the electric theory confines in this case.
The phenomenon of confinement driven by operators
other than mass terms has been observed in other
theories as well \refs{\kut-\ilstr}.

Now we turn to some explicit examples.  To simplify the
ensuing analysis, let us take $y$ to be diagonal with
all non-zero $y_{ii}$ equal.  This breaks the global $SU(N)$
flavor symmetry to an $SU(N-k)\times SO(k)$
symmetry.  (The global $U(1)$'s are also modified; we will
not discuss them here.)
The global symmetries of the magnetic theory include an
$SU(N-k)$ acting on the vectors
and a new flavor group $G_{p,\bar p}$ acting on the spinor representations.
The physics of these theories is quite rich, but for the sake of brevity
we will discuss only three examples.

For example, consider the case $k=2$, for which
$W=\sum_{u=1}^{2} \hat Q^u S \hat Q^u$ and the global
symmetry includes $SU(N-2)\times SO(2)$.  The dual gauge group
is $Spin(6)\approx SU(4)$ with $N-2$ vectors and
two spinors $p,\bar p$ in the ${\bf 4}+{\bf\bar 4}$; its
symmetries include $SU(N-2)$ for the vectors and a
$U(1)\approx SO(2)$ under which $p,\bar p$ have
opposite charge.  The superpotential is $\tilde W = Mqq+Up\bar p$.
The operators are mapped as $Q^{N-4}\to q^2p\overline{p}$,
$\hat Q^u Q^{N-5}\to q^3p^2, q^3\overline{p}^2$,
$\hat Q^2 Q^{N-6}\to q^4p\overline{p}$.
As in \nsewtwo, some operators  have acquired a global
charge as a result of the perturbation.

Consider next the case $k=4$.  The flavor symmetry
of the electric theory is $SU(N-4)\times Spin(4)$.
The magnetic theory has gauge group $Spin(4)\sim SU(2)\times SU(2)$,
with $N-4$ vectors, two spinors $p_a$ in the $({\bf 2},{\bf 1})$,
and two spinors $\bar p_{\dot a}$ in the $({\bf 1},{\bf 2})$
representation.  The flavor-symmetry group of the vectors is $SU(N-4)$
while that of the spinors is $SU(2)\times SU(2)$ in agreement
with the electric theory.  The operators
$Q^{N-4}$, $\tilde QQ^{N-5}$, $\tilde Q^2Q^{N-6}$, $\tilde Q^3Q^{N-7}$,
$\tilde Q^4Q^{N-8}$, which are in the ${\bf 1,4,3+\bar 3,4,1}$ of the
$Spin(4)$ flavor symmetry, are mapped to
$p^2-\overline{p}^2$,  $qp\overline{p}$, $[q^2p^2, q^2\overline{p}^2]$,
$q^3p\overline{p}$, and $q^4(p^2 - \overline{p}^2)$.

As a final example, consider the theory with $N=6$.  This
case is complicated and we do not yet fully understand it.
However, certain aspects of it are under control.  The electric
theory has gauge group $SU(2)$, a triplet $S$ and
six doublets $Q^i$; this is the matter content of an ${\cal N}=2$
supersymmetric theory, but with $W=0$.  Its magnetic dual is
$Spin(8)$ with six vectors and a spinor.
Now consider adding the superpotential
$W=\sum_1^k Q^iSQ^i$.  The dual theory is $Spin(8-k)$
with $6-k$ vectors and the appropriate number of spinors.
In the case $k=6$, the electric $SU(2)$ is an
${\cal N}=2$ supersymmetric theory with three hypermultiplets in the
doublet representation; the flavor symmetry of the theory is
$SO(6)\approx SU(4)$.  This theory and its duality
were studied in \nsewtwo; the electric theory has a Coulomb
branch, parametrized by $U=S^2$, with two singularities.
At one singularity, a dyon becomes massless,
while at the other, monopole hypermultiplets in the
${\bf 4}$ of $SU(4)$ become massless. Here,
when $k=6$, the magnetic theory has a $Spin(2)\approx U(1)$ gauge group,
a neutral field $U$, oppositely charged
fields $p_a$, $\bar p_{\dot a}$, where $a,\dot a=1,\dots 4$, and a
superpotential $W=Up_a\bar p_{\dot a}\delta^{a\dot a}$.
This theory is ${\cal N}=2$ supersymmetric. The point $U=0$, where
$p_a$ and $\bar p_{\dot a}$ are massless,
clearly corresponds to the monopole singularity found by the authors of
\nsewtwo.  The dyon singularity may be identified using
the work of \refs{\sem,\son}. For non-zero $\ev{U}$
of order $\Lambda$, $SU(2)$ with $W=0$ breaks to $SO(2)$ with six doublets
whose magnetic dual is $SO(8)$ with six vectors.
When $W=y_{ij}Q^iSQ^j$, a dyon becomes massless at a point
$U\sim \Lambda^2/\det y $.

Similarly, if we take an $SU(2)$ ${\cal N}=2$ theory
with $N_f<3$ hypermultiplets in the doublet representation,
but we set $W=0$, then the dual theory is $Spin(2N_f+2)$, as can
be derived from above. For $N_f=4$ the magnetic theory is related
to $Spin(10)$ \spinten.

\newsec{$Spin(8)$ and its descendants with superpotential $W=0$}

 {}From the previous sections, a duality
for $Spin(8)$ with $N$ vectors, one spinor, and superpotential $W=0$
follows directly.  To the previous $SU(N-4)$ theory, add singlets
$N_{ij}$, $T$ and the couplings $W = N_{ij}Q^iSQ^j + T\det S$.
To its dual $Spin(8)$ theory one must  add the same singlets, and the
superpotential becomes $\tilde W = NM +TU + Mqq + Upp$.
The singlets are all massive and should be integrated out.
The infrared equations of motion set $M=U=0$, $N=qq$ and $T=pp$;
substituting the equations of motion makes the $Spin(8)$
superpotential vanish, and maps the operators $N$, $T$ in the
$SU(N-4)$ theory to $qq$, $pp$ in $Spin(8)$ while leaving the
rest of the operator mapping \opermap\ unchanged. Thus, $Spin(8)$ with
$q_i$, $p$ and with $W=0$ is dual to $SU(N-4)$ with superpotential
$W = N_{ij}Q^iSQ^j + T\det S$.  All other aspects of the duality
follow {}from this operation.

We will mention only a few features of this model.  In particular
we note that the multiple disjoint branches of $SO(8)$ with
a small number of vector representations emerges correctly
when the spinor of $Spin(8)$ is integrated out. Consider the
non-perturbative structures for small $N$.
For $N<6$ the theory confines and has a dynamically
generated superpotential. For $N=0$, $W=[\Lambda^{17}/T]^{1/5}$, while
for $N=1,2,3$ the superpotential is $[\Lambda^{17-N}/T\det N]^{1/(5-N)}$;
all these are generated by gaugino condensation in the unbroken subgroup
of $Spin(8)$.  These results lead to the correct
$Spin(7)$ dynamical superpotentials \spinseven\
when $\vev{T}\neq 0$.
For $N=4$, instantons generate a superpotential
$T\Lambda^{13}/(T^2\det N+b^2)$; again $\vev{T}\neq 0$ gives
the correct $Spin(7)$ superpotential.    When the spinor
is given mass, the theory develops two physically distinct branches.
Adding $mpp = mT$ to the
superpotential and integrating out the massive fields $b$ and $T$,
one finds the conditions $Tb=0$ and
$(T^2\det N-b^2)\Lambda^{13}/(T^2\det N+b^2)^2 = m$.
On one branch, $T\neq 0$, $b=0$ and the low-energy superpotential
is $W_L\propto\sqrt{\Lambda_L^{14}/\det N}$; on the other branch
$T=0$, $b\neq 0$, there are massless mesons $N_{ij}$, and $W=0$.
This structure of two disjoint branches accords with the results of \son.

The theory with $N=5$ also
confines and has a deformed moduli space given by
$T^2\det N+N_{ij}b^ib^j=T\Lambda^{12}$.    When we add
a mass term for the spinor, the baryons $b^i$ should no longer
be part of the low-energy description.
However, using the chiral anomaly as explained above,
$b_0^{'i} = q^4 W_\alpha^2 = m b^i$, and  the $b^{'i}_0$
are still present in the low-energy theory. The constraint on $T$ implies
\eqn\Tsoln{
T={\Lambda_L^{13}\over m\det N}
\left[1\pm\sqrt{1+{(\det N)N_{ij}b_0^{'i}b_0^{'j}\over \Lambda_L^{26}}}\
\right]}
in terms of the low-energy scale $\Lambda_L^{13}=m\Lambda^{12}$.
For small $\det N$, substituting this
expression in the superpotential gives agreement with \son.
In particular there are two branches, one with $W \sim (\det N)^{-1}$
and one with $W \sim N_{ij}b_0^{'i}b_0^{'j} + \cdots$.

\newsec{Using the Renormalization Group Flow to Check the Duality}

We now consider the renormalization group
flow of a model which flows in one limit to the $SU(N-4)$ theory
studied in this paper, while in another limit it appears
to flow to a different theory.  We will show that, in a highly
non-trivial way, the results of \refs{\sem,\son} and of the
present paper ensure that in the end it flows to the expected
magnetic $Spin(8)$.

Consider the theory $SU(N-4)\times SO(N)$ with fields
$X$, $Q^i$, $i=1,\dots,N$,
in the $({\bf N-4},{\bf N})$, $({\bf\overline{ N-4}},{\bf 1})$
representations.   Let the two gauge groups be characterized
by the scales $\Lambda$, $\Lambda'$.
Suppose that $\Lambda \ll \Lambda'$.  In this case $SO(N)$,
which has $N-4$ vector representations {}from the field $X$, will
confine at the scale $\Lambda'$ without generating
a dynamical superpotential~\refs{\sem,\son}.
The low-energy theory is $SU(N-4)$ with a symmetric
tensor $S\sim XX$ and the $N$ fields $Q^i$; its superpotential
is zero.  This is the theory
under study in this paper, which we have shown to be
dual to $Spin(8)$ with singlets $M^{ij}$, $U$, a spinor
$p$ and vectors $q_i$, and with $W=M^{ij}q_iq_j+Upp$.

Now suppose that $\Lambda \gg \Lambda'$.  This is a physically different
theory {}from the case $\Lambda\ll\Lambda'$, and the two theories
might in principle have different infrared behavior.  However,
we will now show that the theory with $\Lambda\gg\Lambda'$ flows to
the same $Spin(8)$ theory as in the other limit; it does so by
a complicated route,
passing close to three different approximate fixed points
before arriving at its true infrared fixed point.  We view this
result as a strong consistency check on the dualities of
\refs{\sem,\son} together with that of this paper and of \spinseven.

First, the gauge group $SU(N-4)$, which has $N$ flavors {}from $X$
and $Q^i$, becomes strong.  It flows
toward an approximate fixed point (moderately or weakly coupled)
consisting of an $SU(4)$ gauge theory \refs\sem\
times the original $SO(N)$.  The low-energy fields are
$Y^i\sim (XQ^i)$, $\tilde Q_i$, $\tilde X$ in the $({\bf 1},{\bf N})$,
 $({\bf \bar 4},{\bf 1})$, and
$({\bf 4},{\bf N})$ of $SU(4)\times SO(N)$; they are
coupled in the superpotential $W=Y^i\tilde X\tilde Q_i$.

Next, the $SO(N)$ theory, which now has $N+4$
flavors, flows to strong coupling; it flows
to an approximate fixed point with description in
terms of $SU(4)\times SO(8)$ \refs{\sem}.
The fields of this theory are
$\chi \sim (\tilde X\tilde X)$, $\tilde R^i \sim (Y^i\tilde X)$,
$\tilde M^{ij} \sim (Y^iY^j)$, $\tilde Q_i$, $\tilde q_i$, $\tilde x$
in the $({\bf 10},{\bf 1})$,  $({\bf 4},{\bf 1})$, $({\bf 1},{\bf 1})$,
$({\bf \bar 4},{\bf 1})$, $({\bf 1},{\bf 8})$, and
 $({\bf \bar 4},{\bf 8})$ of $SU(4)\times SO(8)$.   Their
superpotential is $W=\tilde R^i(\tilde Q_i+\tilde x q_i)
+ \chi\tilde x\tilde x + \tilde M^{ij}\tilde q_i \tilde q_j$.
Note $\tilde R_i$ and $\tilde Q_i$ are massive and should be integrated
out.

The $SU(4)$ gauge group, which now has a symmetric tensor $\chi$
and eight antifundamental representations $\tilde x$, is pushed away
{}from its approximate fixed point.  {}From this paper, we know that at
strong coupling it is described by a $Spin(8)$ theory with one spinor
and eight vectors.  The flow thus takes this model to
a $Spin(8)\times SO(8)$ description
with gauge singlets $\tilde M^{ij}$, $\tilde U=\det\chi$,
 and charged matter
$Z = \tilde x\chi \tilde x$, $v$, $\tilde q_i$ and $\tilde p$
in the $({\bf 1},{\bf 35_v+1})$, $({\bf 8_v},{\bf 8_v})$,
 $({\bf 1},{\bf 8_v})$, and $({\bf 8_s},{\bf 1})$.
The superpotential is $W = \tr Z + Z(vv) +
\tilde M^{ij}\tilde q_i \tilde q_j + \tilde U\tilde p \tilde p.$

The equation of motion $\del W/\del Z=0$ forces $\tr\vev{vv}\neq 0$.
The D-term conditions then ensure that all $\vev{v_i}$ are equal, and
thus the $Spin(8)\times SO(8)$ theory is broken to the
diagonal $Spin(8)$.  In this process the field $Z$ becomes
massive, and the fields $\tilde M$, $\tilde U$, $\tilde q_i$
and $\tilde p$ remain; the first two are singlets,
$\tilde p$ is a spinor, and the $\tilde q_i$ are vectors of the diagonal
$Spin(8)$.  Their superpotential is
$W = \tilde M^{ij}\tilde q_i \tilde q_j + \tilde U\tilde p \tilde p$,
so we have recovered the dual of the $SU(N-4)$ theory which was found
when $\Lambda\ll \Lambda'$.  The renormalization
group flow is thus self-consistent --- the flow along the two paths
ends at the same $Spin(8)$ fixed point.

\bigskip

\centerline{{\bf Acknowledgments}}

We would like to thank K. Intriligator, D. Kutasov, R. Leigh,
A. Schwimmer and N. Seiberg for useful discussions.
This work is supported in part by DOE grant \#DE-FG05-90ER40559.
P.P. is also supported by a Canadian 1967 Science fellowship.
\listrefs

\bye